\documentclass[prl,aps,twocolumn,nopacs,superscriptaddress,nofootinbib]{revtex4-1}

\usepackage{amsmath}  \usepackage{amssymb}  \usepackage{amsfonts}  \usepackage{bm}  \usepackage{bbm}   \usepackage{bbold}   \usepackage{braket}  \usepackage{color}  \usepackage{comment}  \usepackage{dcolumn}  \usepackage{enumerate}  \usepackage{epsfig}  \usepackage{gensymb}  \usepackage{graphicx}  \usepackage{indentfirst}  \usepackage{lmodern}  \usepackage{mathrsfs}  \usepackage{mathtools}  \usepackage{psfrag}  \usepackage{pst-all}  \usepackage{soul}  

\usepackage{xcolor}

\usepackage{float} 
\usepackage[colorlinks,linkcolor=blue,citecolor=blue,urlcolor=blue,hyperindex,driverfallback=dvipdfm]{hyperref}  \usepackage[T1]{fontenc} 

\def\eepsilon{\epsilon}

\usepackage{dirtytalk}

\def\eepsilon{\epsilon}
\usepackage{CJKutf8}
 
 \usepackage{mathrsfs}
 \usepackage{float} 
\usepackage{placeins}
 \usepackage{bm}
 \usepackage{multicols}
 \usepackage{bm}
 \usepackage[normalem]{ulem} 
 \makeatletter
\newcommand*{\addFileDependency}[1]{
  \typeout{(#1)}
  \@addtofilelist{#1}
  \IfFileExists{#1}{}{\typeout{No file #1.}}
}
\makeatother

\usepackage[strict]{changepage}
\usepackage{calc}
\usepackage{dirtytalk}
\usepackage{pifont}
\newcommand{\xmark}{\ding{55}}
\usepackage{xr-hyper} 
\usepackage{hyperref}

\usepackage{graphicx}
\usepackage[T1]{fontenc}
\usepackage{cancel}
\usepackage{longtable}
\usepackage{array}
\newcolumntype{C}[1]{>{\centering\arraybackslash}p{#1}}
\usepackage{soul}
\usepackage{color}
\usepackage{float}

\newcommand{\ttt}{\mathrm}

\usepackage{mathtools}
\DeclarePairedDelimiter\abs{\lvert}{\rvert}%

\usepackage{ulem}
\usepackage{titlesec}
 \titleformat{\paragraph}[hang]{\bfseries}{}{0pt}{\uline}
\titlespacing*{\paragraph}
{0pt}{3.25ex plus 1ex minus .2ex}{1.5ex plus .2ex}

\usepackage{CJKutf8}

\begin{document}
\begin{CJK*}{UTF8}{gbsn} 

\title{Maximal Violation of Kirchhoff's Law in Planar Heterostructures
}

\author{Lu Wang (汪璐)}
\affiliation{ICFO-Institut de Ciencies Fotoniques, The Barcelona Institute of Science and Technology, 08860 Castelldefels (Barcelona), Spain}

\author{F. Javier Garc\'{\i}a de Abajo  }
\affiliation{ICFO-Institut de Ciencies Fotoniques, The Barcelona Institute of Science and Technology, 08860 Castelldefels (Barcelona), Spain}
\affiliation{ICREA-Instituci\'o Catalana de Recerca i Estudis Avan\c{c}ats, Passeig Llu\'{\i}s Companys 23, 08010 Barcelona, Spain}

\author{Georgia T. Papadakis}
\email{georgia.papadakis@icfo.eu}
\affiliation{ICFO-Institut de Ciencies Fotoniques, The Barcelona Institute of Science and Technology, 08860 Castelldefels (Barcelona), Spain}

\begin{abstract}
Violating Kirchhoff's law has so far required nonreciprocal materials patterned in microstructures. In these configurations, the excitation of a guided or polaritonic mode that lies outside the light cone, often via gratings, was a requirement. Here, we describe how nonreciprocity manifests itself in \textit{pattern-free} heterostructures. We demonstrate that a resonant mode in a dielectric spacer separating a nonreciprocal film from a back-reflector suffices to maximally violate Kirchhoff's law, and identify the minimal dielectric requirements for such functionality, which are satisfied by currently available materials.
\end{abstract}

\date{\today}

\maketitle
Global energy demands call for renewable energy production at the terawatt scale and beyond \cite{xu2018global}. Light-harvesting renewable energy approaches, such as solar photovoltaic cells, can reach a performance near thermodynamic limits if the fundamental constraint of Kirchhoff's law of thermal radiation is broken \cite{green2012time,park2022nonreciprocal}. Kirchhoff's law states that a material's absorptivity $\alpha$ ought to equal its thermal emissivity $e$ for every frequency and direction. By violating Kirchhoff's law, one can efficiently redirect emitted photons from one energy converter to another in a concatenated energy-conversion scheme, leading to an ultimate energy conversion efficiency of $93\%$ (Landsberg's limit) \cite{landsberg1980thermodynamic}. So far, several photovoltaic configurations have been proposed, operating both in reflection \cite{green2012time} and transmission geometries \cite{park2021reaching,park2021violating}. 

Fundamentally, breaking Kirchhoff's law of thermal radiation requires nonreciprocal materials that break time-reversal symmetry. This is often realized by applying an external magnetic field to magneto-optical materials, such as InAs \cite{caloz2018electromagnetic,buddhiraju2020nonreciprocal,zhu2014near}. Nevertheless, high, tesla-scale external magnetic field strengths are typically required \cite{shayegan2022nonreciprocal,figotin2001nonreciprocal,zhu2014near,zhao2019near}, thus resulting in structures that are bulky, expensive, and unsuited to large-scale manufacturing \cite{kord2020microwave}. Hence, the first experimental realization of nonreciprocal emission at mid-infrared (mid-IR) frequencies was reported just last year \cite{shayegan2022nonreciprocal}. In that work, a guided resonant mode was excited in the Voigt configuration via a grating \cite{shayegan2022nonreciprocal}. This result followed several similar theoretical proposals \cite{zhao2019near} that considered the excitation of a guided-mode resonance in order to amplify the intrinsic nonreciprocal material response.

To alleviate the requirement of high-magnetic fields, magnet-free nonreciprocal materials, namely Weyl semimetals, have been recently explored \cite{zhao2020axion,wang2021broadband,wu2021near}. This emerging class of quantum materials possesses unique topological properties, leading to magnetic-like effects even in the absence of an external magnetic field \cite{jia2016weyl,kotov2018giant}. 
Several Weyl semimetals have been already experimentally identified, such as $\text{WP}_\text{2}$, $\text{Y}_2\text{Ir}_2\text{O}_7$, $\text{HgCr}_2\ttt{Se}_4$, $\ttt{TaAs}$, and $\text{Co}_3\ttt{Sn}_2\ttt{S}_2$  \cite{kumar2017extremely,wan2011topological,xu2011chern,yan2017topological,okamura2020giant}. So far, Weyl materials have been considered by the photonics community as candidates for nonreciprocal thermal emission in theoretical proposals involving geometries such as gratings \cite{zhao2020axion,park2021violating}, photonic crystals \cite{li2021strong}, and prisms \cite{wu2021enhanced}. In the majority of works, similar to previous studies with magneto-optical materials, a resonant guided mode \cite{park2021violating,shayegan2022nonreciprocal} or a polaritonic mode \cite{zhao2020axion} is excited in the Voigt configuration \cite{budker2002resonant}.
\begin{figure*}[th]
\centering
\includegraphics[width=1\linewidth]{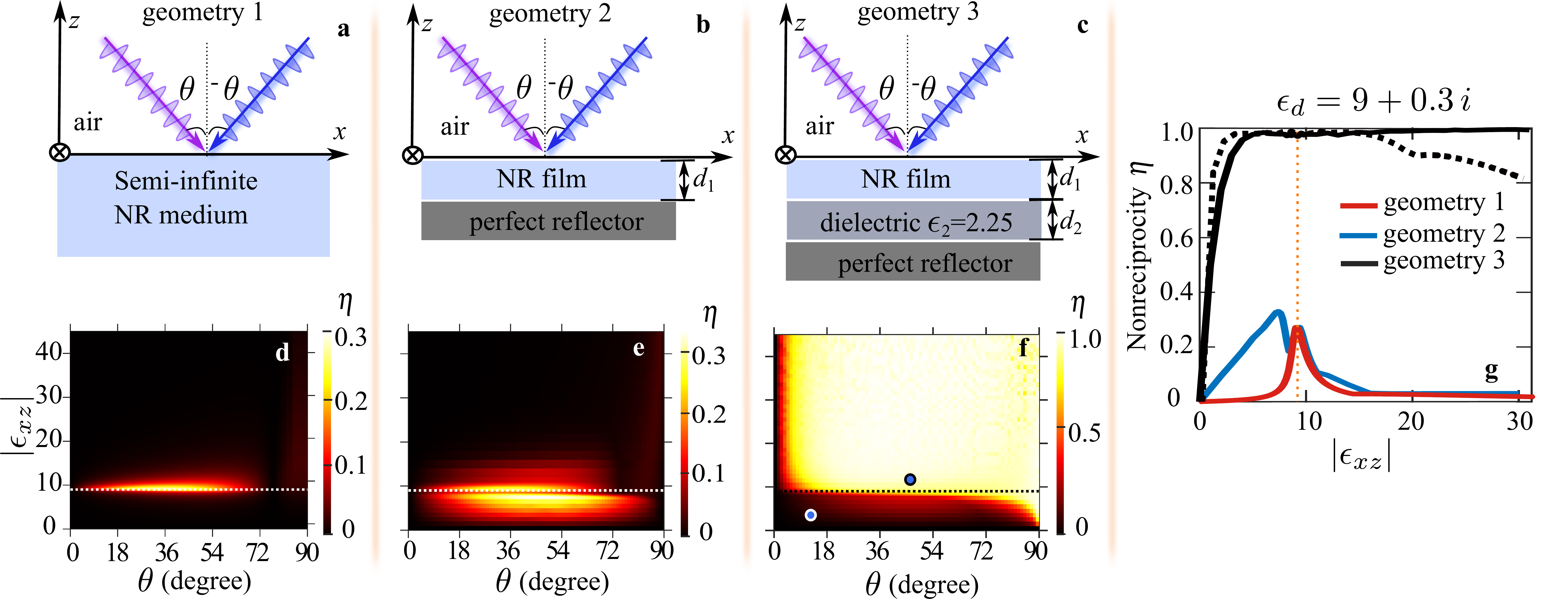}
\caption{Nonreciprocity in simple planar structures: \textbf{a} a semi-infinite surface, \textbf{b} a NR film of thickness $d_1$ on a back-reflector, and \textbf{c} a dielectric spacer separating a NR film from a back-reflector. Color plots in \textbf{d}-\textbf{f} show the maximum nonreciprocity $\eta$ within a wide range of thicknesses $d_\mathrm{1}$ and $d_\mathrm{2}$ for each geometry as a function of the off-diagonal tensor element of the nonreciprocal material ($\epsilon_{xz}$) and the angle of incidence $\theta$.  Without loss of generality, we set $\epsilon_d=9+0.3\,i$. The horizontal and vertical dashed lines indicate $\abs{\epsilon_{xz}}={\rm Re}\{\epsilon_d\}$. For geometry 3 and panels \textbf{c}, \textbf{f}, we set $\epsilon_2=2.25$. \textbf{g} Optimum $\eta$ value in the maps of panels \textbf{d-f} as a function of $\abs{\epsilon_{xz}}$. Solid curves represent results calculated with $\epsilon_d=9+0.3\,i$, whereas the dashed curve corresponds to $\epsilon_d=-9+0.3\,i$ for geometry 3 (see more details in supplementary material S4.A).}
\label{fig:1}
\end{figure*}

Both material classes, magneto-optical materials, and Weyl semimetals, are described via their dielectric permittivity tensor. In the Voigt configuration \cite{zhao2020axion}, this tensor takes the form
\begin{align}
\bm{\epsilon}_{\rm NR}=
    \begin{pmatrix}
        \epsilon_{d} & 0 &  \epsilon_{xz}\\
        0 & \epsilon_{d}& 0   \\
        -\epsilon_{xz} & 0 &  \epsilon_{d}
    \end{pmatrix}\label{eq:eps_voigt},
\end{align}
where $\epsilon_{d} \in \mathbb{C}$ and $\epsilon_{xz}=i\epsilon_\mathrm{a}$ with $\epsilon_\mathrm{a} \in \mathbb{R}$. For simplicity, we assume that the diagonal tensor elements are all equal. The nonzero, imaginary $\epsilon_{xz}$ component results from an applied magnetic field (Weyl nodes separation) in magneto-optical materials (Weyl semimetals) along the $y$ direction. A violation of Kirchhoff's law reflects into a nonzero value of $\eta=\abs{e(\theta)-\alpha(\theta)}$. Henceforth, the parameter $\eta$ is termed nonreciprocity.

The number of nonzero tensor elements in the description of Eq. (\ref{eq:eps_voigt}) makes the analytical description of nonreciprocal materials in the aforementioned inhomogeneous nano- and micro-structures rather challenging. Here, in contrast to previous works relying heavily on numerical solvers, we derive simple analytical equations that describe how nonreciprocity manifests itself in planar, pattern-free geometries. We show theoretically that planar structures can maximally violate Kirchhoff's law, approaching the limit $\eta=1$ over a wide range of incidence angles. We show that the considered configurations do not require the excitation of guided modes that lie outside the light cone. In contrast, the near complete violation of Kirchhoff's law stems from wave interference as in conventional resonant absorption devices \cite{unlu1995resonant}. Finally, we numerically identify the minimum requirement for the off-diagonal permittivity tensor element $\epsilon_{xz}$ to produce a maximal violation of Kirchhoff's law, and we classify currently available nonreciprocal materials in terms of this parameter.

We start by considering three standard planar geometries as shown in Fig.~\ref{fig:1}: a semi-infinite nonreciprocal medium (panel \textbf{a}), a nonreciprocal layer on a back-reflector (panel \textbf{b}), and a dielectric spacer separating a nonreciprocal film from a back-reflector (panel \textbf{c}). For these  non-transmissive geometries, the nonreciprocity $\eta$ can be simplified to $\eta=\abs{e(\theta)-\alpha(\theta)}=\abs{R(\theta)-R(-\theta)}$, which is equivalent to $\abs{R(k_x)-R(-k_x)}$ \cite{zhu2014near,zhang2020validity,wu2022dual}, where $\theta$ is the angle of incidence as shown in Figs.~\ref{fig:1}(\textbf{a}-\textbf{c}), and $k_x$ is the wave vector along $x$ (for details see Sec.~S2 in SM \cite{EPAPSnonreciprocal}). The reflectance is defined as $R=\abs{r}^2$, where $r$ is the reflection coefficient. In particular, in Fig.~\ref{fig:1}, the diagonal tensor element $\epsilon_d=9+0.3\,i$ is chosen as an example.

The normal to the interfaces is aligned with the $z$-axis. The layers of air, nonreciprocal material, and lossless dielectric, respectively, are labeled with subscripts 0-2. For example, $k_{z0}=\sqrt{k_0^2-k_x^2}$ is the wave vector along $z$ in the air. From momentum conservation, $k_x$ remains the same in all the layers. In the bulk of the nonreciprocal material, one obtains four solutions for $k_z$, where two correspond to $s$-polarized fields and two for $p$-polarized fields (see Sec.\ S1 of SM \cite{EPAPSnonreciprocal}).

With the choice of the permittivity tensor of Eq.~(\ref{eq:eps_voigt}), $s$- and $p$-polarized electric fields are decoupled.
In addition, the $s$-polarized fields do not experience any nonreciprocal response (see Sec.~S1 in SM \cite{EPAPSnonreciprocal}). Thus, we focus here on $p$-polarized fields defined as $\mathbf{E}( \mathbf{r},t)=2{\rm Re}\{(E_x,0,E_z)\exp{[i(\mathbf{k}\cdot \mathbf{r}-\omega t)]}\}$ in the air, where $\mathbf{k}=(k_x,0,k_z)$. It is then convenient to define $\eepsilon_{v}=\epsilon_{d}-\abs{\epsilon_{xz}}^2/\epsilon_{d}$, such that the z-component of the wavevector in the nonreciprocal material is written as $k_{z1}=\sqrt{\epsilon_{v}k_0^2-k_x^2}$.

First, we treat the problem fully analytically and derive the reflection coefficient for geometry 1 (see Secs.~S1 and S3 in SM \cite{EPAPSnonreciprocal}).
\begin{align}
r=\frac{k_{z0}\eepsilon_{v}-k_{z1}+k_x\eepsilon_{xz}/\eepsilon_{d}}{k_{z0}\eepsilon_{v}+k_{z1}-k_x\eepsilon_{xz}/\eepsilon_{d}}\label{eq:rlayer}.
\end{align}
 It is clear from Eq.~(\ref{eq:rlayer}) that  $R(k_x)\equiv\abs{r(k_x)}^2\neq R(-k_x)$, even for the semi-infinite interface formed by the air and nonreciprocal material. Thus, this geometry in principle suffices to induce a nonreciprocal effect without spatial patterning or coupling to the surface or guided modes. We note that by setting the denominator of Eq.~(\ref{eq:rlayer}) to zero, we obtain the surface plasmon dispersion relation $k_{z0}\eepsilon_{v}+k_{z1}-k_x\eepsilon_{xz}/\eepsilon_{d}=0$ \cite{zhao2020axion,hu2015surface,chiu1972magnetoplasma}, whereas by setting $\eepsilon_{xz}=0$, Eq.~(\ref{eq:rlayer}) reduces to the standard (reciprocal) Fresnel reflection coefficient for $p$ polarization. In Fig.~\ref{fig:1}\textbf{(d)}, we show that there is an optimal value of $\eepsilon_{xz}$ for maximal violation of Kirchhoff's law. Counter-intuitively, a larger $\eepsilon_{a}$ does not necessarily lead to stronger nonreciprocity. In fact, for very large $\eepsilon_{xz}$, $R(k_x)$ and $R(-k_x)$ both approach unity, thus the nonreciprocity  $\eta=\abs{R(k_x)-R(-k_x)}$ vanishes. The same conclusion holds in geometries 2 and 3 (i.e $\eta\to0$ when $\abs{\eepsilon_{xz}}\to\infty$), though the values of $\abs{\eepsilon_{xz}}$ explored in Fig.~\ref{fig:1}(\textbf{f}) are not large enough to clearly observe this trend.

The reflection coefficient for geometry 2 is given by
\begin{align}
& r=
\frac{\eepsilon_{xz}\,k_x+i\eepsilon_d\,k_{z1}/\tau_1 -(\eepsilon_d\,k_0^2-k_x^2)/k_{z0}}{\eepsilon_{xz}\,k_x +i\eepsilon_d\,k_{z1}/\tau_1  +(\eepsilon_d\,k_0^2-k_x^2)/k_{z0}}\label{eq:r2layer}
\end{align}
with $\tau_1=\tan(k_{z1}d_1)$ and $d_1$ is the thickness of the nonreciprocal material. Figure~\ref{fig:1}(\textbf{e}) indicates that the nonreciprocity, $\eta$ differs significantly from that of geometry 1. Furthermore, geometry 2 lacks tunability, because the absorption and phase of the reflected fields solely depend on the thickness of the nonreciprocal material slab. 

To enhance the tunability in the design of the heterostructure, we insert a lossless dielectric layer with thickness $d_2$ below the nonreciprocal material, as shown by Fig.~\ref{fig:1}(\textbf{c}). This extra layer introduces an additional degree of freedom in optimizing the nonreciprocal thermal emitter, via imposing a tunable phase to the reflected fields. Thus, due to interference, the reflected fields, as well as the parameter $\eta$, are both strongly dependent on  $d_2$. We analyze geometry 3 through the expression
\begin{widetext}
\begin{align} \label{eq:r3layer}
r=-\frac{
(i \eepsilon_{xz} \tau_2 k_{z2} -\eepsilon_2 k_x) (k_x+\eepsilon_{xz}k_{z 0}) /\eepsilon_d
+k_{z1} (\tau_2 k_{z2} -i \eepsilon_2 k_{z0}) /\tau_1
+(\eepsilon_2 k_0^2 +i \eepsilon_d \tau_2 k_{z2} k_{z 0})
}{
(i \eepsilon_{xz} \tau_2 k_{z2}-\eepsilon_2 k_x ) (k_x-\eepsilon_{xz} k_{z 0}) /\eepsilon_d
+k_{z1} (\tau_2 k_{z2} +i \eepsilon_2 k_{z0}) /\tau_1
+(\eepsilon_2 k_0^2 -i \eepsilon_d \tau_2 k_{z2} k_{z 0})
}
\end{align}
\end{widetext}
with $\tau_2=\tan(k_{z2}d_2)$. We note that Eq. (\ref{eq:r3layer}) converges to Eq.~(\ref{eq:r2layer}) when setting $d_2=0$, and this in turn to Eq.~(\ref{eq:rlayer}) by taking the limit $\lim d_1\to\infty$. Further, we note that $\eta$ vanishes if $\theta=0$ or if $\eepsilon_{xz}=0$, as expected. Additionally, $\eta=0$ when ${\rm Im}\{\eepsilon_d\}$ is zero [see sec.~S4 in SM \cite{EPAPSnonreciprocal}]. This is also expected as lack of optical loss prohibits thermal emission from the fluctuation-dissipation theorem \cite{mandel1995optical}. As a rule of thumb, the maximal violation of Kirchhoff's law ($\eta\sim 1$) can be obtained via the condition $\abs{{\rm Re}\{\eepsilon_d\}}\leq \abs{\eepsilon_{xz}}$, provided that $\abs{\eepsilon_{xz}}$ is not too large as discussed above.

A nonreciprocity enhancement is observed over a broad range of incidence angles and off-diagonal permittivity values, as shown in Fig.~\ref{fig:1}(\textbf{f}). In Fig.~\ref{fig:1}(\textbf{g}), we demonstrate explicitly the dependence of $\eta$ on $|\eepsilon_{xz}|$ via selecting the maximum among all incidence angles in panels \textbf{d-f}. We find that, for geometry 3, $\eta$ approaches unity for considerably smaller values of $|\eepsilon_{xz}|$ as compared to geometries 1 and 2. Thus, in practice, due to the typically small values of $|\eepsilon_{xz}|$ that are available, for example in magneto-optical materials (see Table \ref{tab:parameters}), the three-layer geometry 3 is favorable.

\begin{figure}[ht]
  \includegraphics[width=0.75\linewidth]{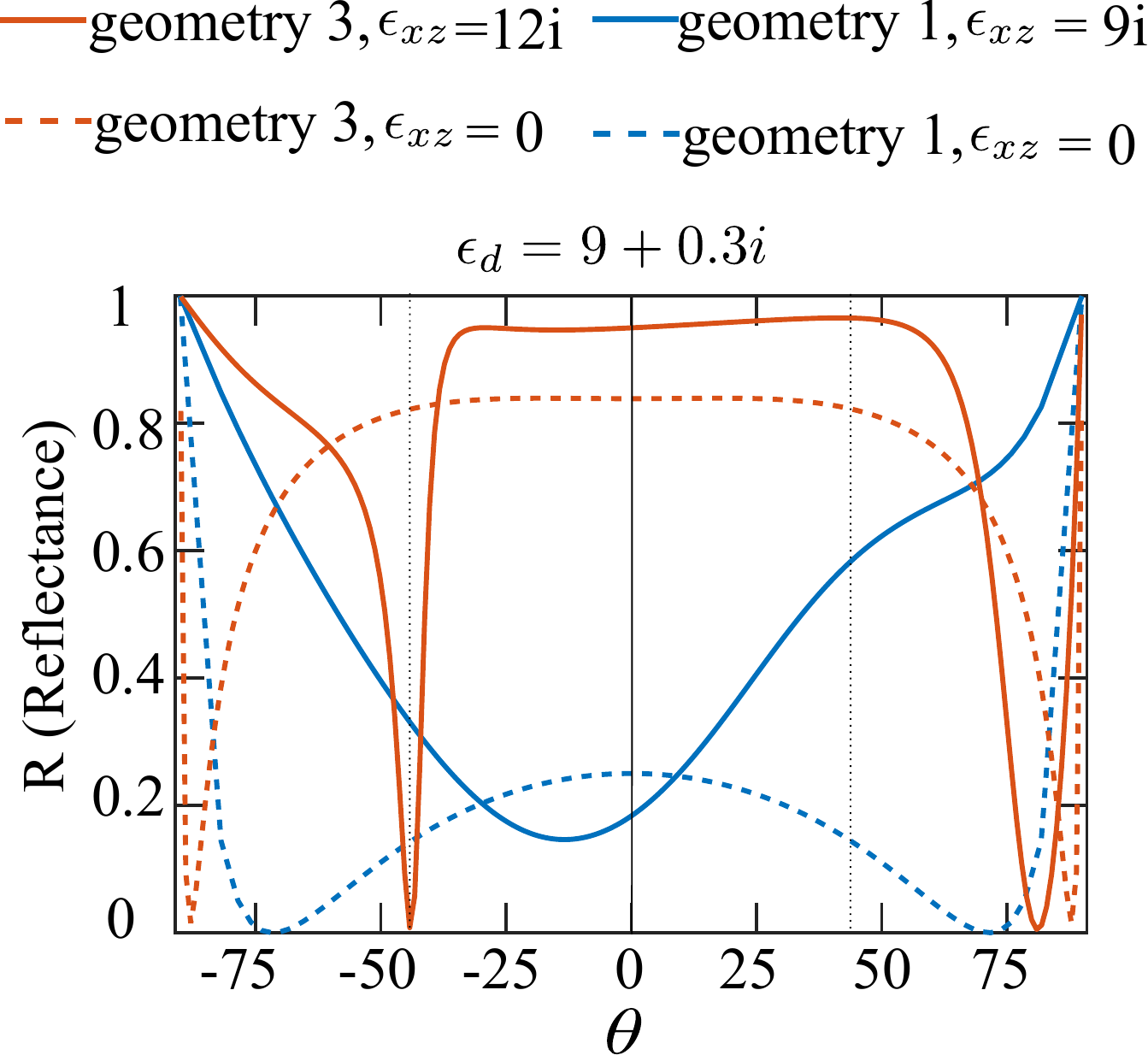}
\caption{Reflectance as a function of incidence angle $\theta$ for reciprocal ($\epsilon_{xz}=0$, dashed curves) and nonreciprocal (solid curves, $\epsilon_{xz}\neq 0$) structures. Geometry 1 (blue curves) and geometry 3 (red curves) are chosen as examples. In particular, for geometry 3, the solid (dashed) curve corresponds to thicknesses  $d_1/\lambda_0=0.10, d_2/\lambda_0=0.32$ 
($d_1/\lambda_0=0.30,d_2/\lambda_0=0.11$). In  the red solid curve, a maximum nonreciprocity $\eta=0.96$ is obtained at $\theta=\pm  43^ \circ$ (marked by vertical dashed lines; see Fig.~\ref{fig:field_profile} for details).  }\label{fig:reflection}
\vspace{-0.7cm}
\end{figure} 
In Fig.\ref{fig:reflection}, we show the reflectance as a function of the incidence angle for reciprocal and nonreciprocal instances of geometries 1 and 3, respectively. We conclude that the nonreciprocity is significantly enhanced through the addition of a dielectric spacer of optical thickness $d_2/\lambda_2<1$. Besides, a nonzero $\epsilon_{xz}$ leads to asymmetric reflection with respect to the incident angle $\theta$ for both geometries. In geometry 3, the nonreciprocity exhibits an asymmetric resonant response as a function of $\theta$, which leads to stronger nonreciprocal effects in the optimal selection of $|\eepsilon_{xz}|$ with respect to geometry 1, as shown in Fig.~\ref{fig:1}(\textbf{g}).

Based on the analytical expressions in Eqs. (\ref{eq:rlayer}-\ref{eq:r3layer}) for the reflection coefficient in the presence of nonreciprocity for the planar geometries 1-3, we can derive design rules for nonreciprocal thermal emitters. In particular, in Fig.~\ref{fig:1}(\textbf{g}), we have demonstrated that geometry 3 requires a smaller value of $|\eepsilon_{xz}|$ for achieving the same level of nonreciprocal response ($\eta$). Thus, henceforth, we focus on geometry 3 and evaluate in more detail Eq. (\ref{eq:r3layer}).

\begin{table*}[ht!]
\centering
\begin{tabular}{l l|c|c|c|c}
    \hline
    material & & ${\rm Re}\{\epsilon_d\}$ & ${\rm Im}\{\epsilon_d\}$ & $\abs{\epsilon_{xy}}$ & B (T)\\ 
    \hline
    $\text{Eu}_2\text{Ir}\text{O}_7$ & \cite{zhao2020axion,kotov2018giant}   & -25 --- 5 &2& 4 --- 20 & \xmark\\
    Co$_3$Sn$_2$S$_2$ (20 K) & \cite{okamura2020giant,xu2020electronic}   & 20 --- 40 & 20 --- 40 &  25 --- 45& \xmark\\
    WSM standard (0 K) & \cite{chen2019optical} & -10 --- 40& 5 --- 30 & 5 --- 25& \xmark \\
    InAs & \cite{shayegan2022nonreciprocal}   &3.6 --- 10.9  & 2.1$\times10^{-2}$ --- 3.4$\times10^{-1}$& 7.5$\times10^{-2}$ --- 1.2 & 3\\
    GaAs & \cite{wang2018nonreciprocal}   & 6.6 --- 7.8 & 7.5$\times10^{-3}$ --- 1.2$\times10^{-1}$& 1.0$\times10^{-2}$ --- 1.7$\times10^{-1}$  & 3\\
    \hline
\end{tabular}
\caption{Parameter ranges for several available magneto-optical materials [see Eq.~(\ref{eq:eps_voigt})] within the thermal emission wavelength range of $6-15\,${\textmu}m: InAs and GaAs under a magnetic field of 3$\,$T, along with several Weyl semimetals in the absence of a magnetic field [marked with \xmark ~in the B(T) column]. Reported values correspond to room temperature unless otherwise stated.}
\label{tab:parameters}
\end{table*}

We obtain results using the transfer-matrix method \cite{mackay2020transfer,berreman1972optics,passler2017generalized,yeh1979electromagnetic} (see SM \cite{EPAPSnonreciprocal} Sec.~S1). Since the nonreciprocity $\eta$ is a periodic function of $d_1$ and $d_2$ [Eq. (\ref{eq:r3layer})], we consider $d_1/\lambda_1$ and $d_2/\lambda_2$ in the range $0-3$, as this range contains the sought-after maxima of $\eta$, where $\lambda_1=\text{max}\big({\rm Re}\{\lambda_0/\sqrt{\eepsilon_v}\},\,{\rm Im}\{\lambda_0/\sqrt{\eepsilon_v}\}\big)$, $\lambda_2=\lambda_0/\sqrt{\eepsilon_2}$. Furthermore, we consider ranges of $\epsilon_{d}$ and $|\eepsilon_{xz}|$ that correspond to known magneto-optical materials and Weyl semimetals within the thermal wavelength region of $6-15\,${\textmu}m at room temperature, as shown in Table \ref{tab:parameters}. In particular, we choose the calculation parameter ranges ${\rm Im}\{\eepsilon_{d}\} \in\{0.3,3,30\}$, ${\rm Re}\{\eepsilon_{d}\}  \in[-35,35]$, and  ${\rm Im}\,\{\epsilon_{xz}\} \in[0,40]$. We note that, although only positive values of ${\rm Im}\,\{\epsilon_{xz}\}$ are discussed here, negative ${\rm Im}\,\{\epsilon_{xz}\}$ values are automatically accounted for because this is equivalent to rotating $\bm{\epsilon}_{\rm NR}$ in Eq.~(\ref{eq:eps_voigt}) by $180^\circ$ around the $z$ axis, which leads to invariant results by simultaneously flipping the signs of $\eepsilon_{xz}$ and $k_x$.

\begin{figure}[ht]
\centering
  \includegraphics[width=1\linewidth]{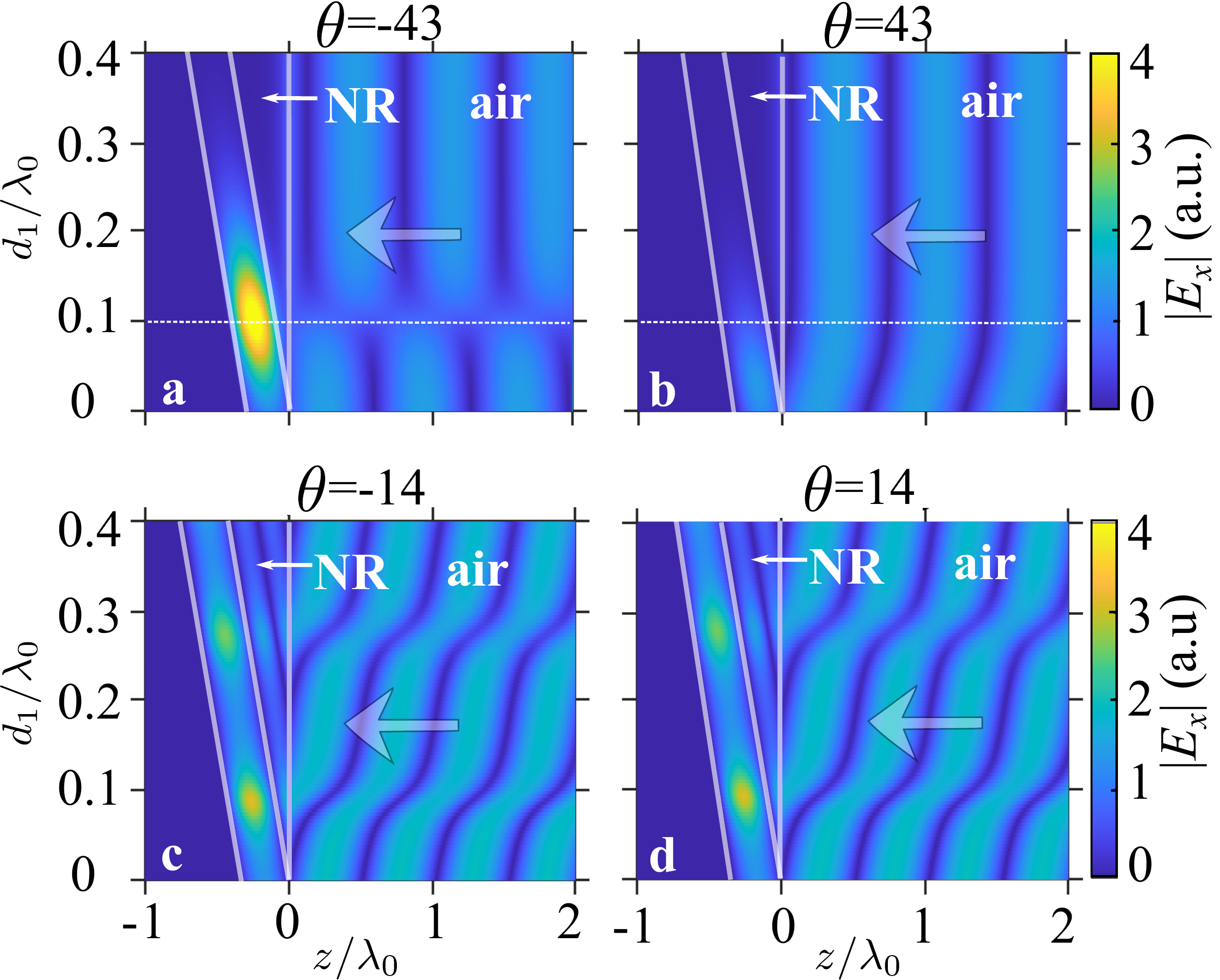}
\caption{Field profiles corresponding to the conditions marked by blue dots in Fig.~\ref{fig:1}\textbf{(f)} (geometry 3 with $\epsilon_d=9+0.3\,i$). We plot the $x$-field amplitude as a function of the normalized distance $z/\lambda_0$ relative to the air/NR interface (horizontal axes) and normalized NR film thickness $d_1/\lambda_0$ (vertical axes). Thin white solid vertical lines indicate the interfaces in the planar structures. The white transparent arrow represents the field incidence direction. (\textbf{a},\textbf{b}) correspond to a nonreciprocity $\eta=0.96$ with incidence angle $\theta=\pm  43^ \circ$ and a dielectric spacer thickness $d_2/\lambda_0=0.32$. The vertical dashed line corresponds to the exact thickness selection in Fig.2. (\textbf{c},\textbf{d})  correspond to  $\eta=0.05$ with $\theta=\pm 14 ^ \circ$ and $d_2/\lambda_0=0.34$. }\label{fig:field_profile}
\end{figure}

\begin{figure}[ht]
  \includegraphics[width=1\linewidth]{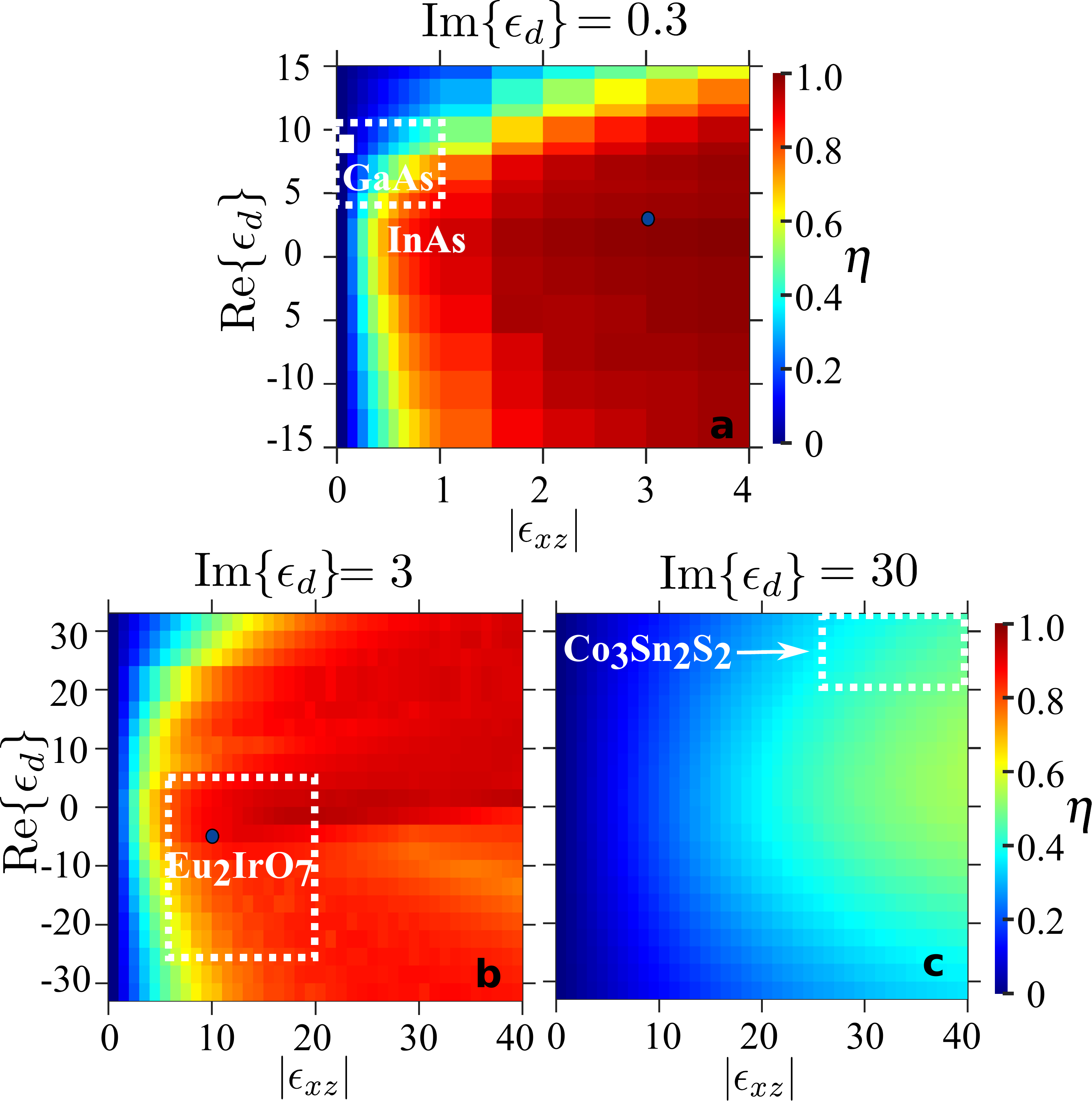}
\caption{Maximum nonreciprocity $\eta$ for geometry 3 [Fig.~\ref{fig:1}(\textbf{c})] as a function of the absolute value of the off-diagonal permittivity $\abs{\epsilon_{xz}}$ and the real part of the diagonal permittivity ${\rm Re}\{\eepsilon_d\}$. Each data point is set to the maximum within the parameter space $d_1$, $d_2$, and $\theta$ defined in the main text. In particular, we show two examples (marked by the blue dots in panels \textbf{a,b}) of $d_1$, $d_2$, and $\theta$ dependence in SM Fig.S5. 
 \cite{EPAPSnonreciprocal} }\label{fig:epsi}
\end{figure}

To understand the origin of the strong nonreciprocal response of geometry 3, in Fig.~\ref{fig:field_profile}, we present the field profiles supported in this geometry for minimal and maximal violation of Kirchhoff's law, corresponding to $\eta=0.05$ and $\eta=0.96$, respectively. These values of nonreciprocity are represented in Fig.~\ref{fig:1}(\textbf{f}) by two blue dots. The parameter $\eta$ is a periodic function of $d_1/\lambda_0$ and $d_2/\lambda_0$. In Fig.~\ref{fig:field_profile} we consider values of $d_1/\lambda_0$ (vertical axes) varying from 0 to 0.4, including the first maximum of $\eta$. In particular, Figs.~\ref{fig:field_profile}(\textbf{a},\textbf{b}) show results for $\eepsilon_{xz}=12\,i$, $d_2/\lambda_0=0.32$, and $\theta=\pm 43^ \circ $, where $+$ and $-$ correspond to $k_x>0$ and $k_x<0$, respectively. Figures~\ref{fig:field_profile}(\textbf{c},\textbf{d}) show field profiles calculated for $\eepsilon_{xz}=4\,i$, $d_2/\lambda_0=0.34$, and $\theta=\pm 14^ \circ$. As is evident in Fig. \ref{fig:field_profile}, a maximal violation of Kirchhoff's law occurs when the field intensity is maximum inside the dielectric spacer. In other words, $\eta \sim 1$ is associated with a resonant mode inside the dielectric spacer, excited when light is incident from one side ($-k_{x}$), while it is suppressed when incident from the other side ($+k_{x}$). 

Last, we analyze how material losses affect nonreciprocity by exploring three distinct values of ${\rm Im}\{\eepsilon_d\}\in \{0.3,3,30\}$, which span the entire parameter range in Table \ref{tab:parameters}, corresponding to realistic materials. In Fig.~\ref{fig:epsi}, we compute $\eta$ when ${\rm Re}\{\eepsilon_d\}$ and $\eepsilon_{xz}$ are varied. The parameter ranges of standard magneto-optical materials (GaAs and InAs) are shown in panel \textbf{a}, whereas material properties representing Weyl semimetals Eu$_2$IrO$_7$ and Co$_3$Sn$_2$S$_2$ are shown in panels \textbf{b} and \textbf{c}, respectively.  For large ${\rm Im}\{\eepsilon_d\}$, the maximum of nonreciprocity tends to shift closer to $\theta\sim 90^ \circ$ (i.e., grazing incidence, see Fig.~S4 in SM \cite{EPAPSnonreciprocal}). The white contours in these figures represent the range of values of each labeled material as reported in recent literature. From these figures, we have identified the degree of nonreciprocal response that each considered material can reach, upon optimizing $d_1$ and $d_2$.

In conclusion, we have shown that, even in a planar, pattern-free heterostructure, one can maximally violate Kirchhoff's law of thermal radiation. This can occur without the excitation of guided or polaritonic modes that lie outside the light cone. We described analytically how the off-diagonal tensor element in the permittivity  of a nonreciprocal material ($\epsilon_{xz}$) manifests itself in the reflection from a planar structure. Based on our theory, we show that the requirement for large values of $\epsilon_{xz}$ is relaxed in a three-layered geometry consisting of a nonreciprocal material on a dielectric spacer on a back-reflector, whereas in a semi-infinite nonreciprocal material, a large value of $\epsilon_{xz}$ does not necessarily lead to stronger nonreciprocity. Our theory is general and applies to both magneto-optical materials and Weyl semimetals, which we have classified in our work in terms of nonreciprocal thermal emission performance. Our analysis may serve to identify design rules for simpler nonreciprocal thermal emitters.


We thank Prof.~Mehrdad Shokooh-Saremi,  Prof.~Bo Zhao, and Dr.~\'Alvaro Rodr\'{\i}guez Echarri for stimulating discussions. L.~W.~thanks Mr.~Juli Céspedes and  Mr.~Muhammad-Zeshan Sayab for IT-related discussions.
We acknowledge funding from the ecological transition grant, European Union [Horizon 2020 Marie Sklodowska-Curie grant No.~847648 (fellowship code LCF/BQ/PI21/11830019)], \say{la Caixa} Foundation (ID 100010434), the Spanish MICINN (CEX2019-000910-S from AEI/10.13039/501100011033, PID2020–112625 GB-I00, and Severo Ochoa CEX2019-000910-S), the Generalitat de Catalunya (CERCA, AGAUR), TED2021-129841A-I00 funded by MCIN/AEI/ 10.13039/501100011033 and by the European Union “NextGenerationEU”/PRTR, and Fundaci\'os Cellex and Mir-Puig. L.W. acknowledges support from the Severo Ochoa post-doctoral fellowship. 


\end{CJK*}
\bibliography{pic_bib/apssamp} %

\end{document}